\documentclass[preprint,12pt]{article}

\usepackage{amssymb}

\usepackage{amsmath}

\usepackage{hyperref}

\begin{document}





\centerline{}

\centerline {\Large{\bf The exactly solvable two - dimensional }}

\centerline{}

\centerline{\Large{\bf stationary Schr\"odinger operators
obtaining }}

\centerline{}

\centerline{\Large{\bf by the nonlocal Darboux transformation}}

\centerline{}

\centerline{\bf {A. G. Kudryavtsev}}

\centerline{}

\centerline{Institute of Applied Mechanics,}

\centerline{Russian Academy of Sciences, Moscow 119991, Russia}

\centerline{kudryavtsev\_a\_g@mail.ru}

\begin{abstract}

The Fokker-Planck equation associated with the two - dimensional
stationary Schr\"odinger equation has the conservation low form
that yields a pair of potential equations. The special form of
Darboux transformation of the potential equations system is
considered. As the potential variable is a nonlocal variable for
the Schr\"odinger equation that provides the nonlocal Darboux
transformation for the Schr\"odinger equation. This nonlocal
transformation is applied for obtaining of the exactly solvable
two - dimensional stationary Schr\"odinger equations. The examples
of exactly solvable two - dimensional stationary Schr\"odinger
operators with smooth potentials decaying at infinity are
obtained.

\end{abstract}






\section{Introduction}

Consider the two - dimensional stationary Schr\"odinger equation
\begin{equation} \label{eq1}
{\frac {\partial ^{2}}{\partial {x}^{2}}}Y \left( x,y \right)
+{\frac {\partial ^{2}}{\partial {y}^{2}}}Y \left( x,y \right)
-{\it u}
 \left( x,y \right) Y \left( x,y \right)
 =0
\end{equation}

In the case ${\it u}=-E+V\left( x,y \right)$ equation \eqref{eq1}
describes nonrelativistic quantum system with energy $E$ \cite
{Landau}. In the case ${\it u}= {\omega} ^{2}/{c\left( x,y
\right)}^{2}$ equation \eqref{eq1} describes an acoustic pressure
field with temporal frequency $\omega$ in inhomogeneous media with
sound velocity $c$ \cite {Morse}. The case of fixed energy $E$ for
two - dimensional equation is of interest for the multidimentional
inverse scattering theory \cite {Grinevich2000} due to connections
with two - dimensional integrable nonlinear systems \cite
{Veselov1984}, \cite {Novikov2010}. The case of fixed frequency
$\omega$ is of interest for modelling in acoustic tomography \cite
{Kak}.

The useful tool for one - dimensional Schr\"odinger equation is
the Darboux transformation \cite {Matveev1991}. Straightforward
generalizations of Darboux transformation for two - dimensional
case were proposed including  operators of second order in
derivatives \cite {Andrianov1995}, \cite {Ioffe2004} but set of
exactly solvable two - dimensional models obtained is rather
limited. Some examples of exactly solvable two - dimensional
stationary Schr\"odinger operators with smooth rational potentials
decaying at infinity were obtained in the papers \cite
{Tsarev2007}, \cite {Tsarev2008} by application of the Moutard
transformation which is a two-dimensional generalization of the
Darboux transformation \cite {Tsarev2010}. In the past years
progress was made in the symmetry group analysis of differential
equations by extending the spaces of symmetries of a given partial
differential equations system to include nonlocal symmetries \cite
{Ibragimov1991}, \cite {Bluman2010}. In the present paper the
nonlocal variable is included in Darboux transformation for
investigation of exactly solvable two - dimensional stationary
Schr\"odinger equations.

\section{The nonlocal Darboux
transformation for the Schr\"odinger equation}

Substituting the following expression into equation \eqref{eq1}

\begin{equation} \label{eq2}
Y \left( x,y \right) =W \left( x,y \right) {e^{h \left( x,y
\right) }}
\end{equation}

we obtain the Fokker-Planck equation

\begin{equation} \label{eq3}
{\it W_{xx}}+{\it W_{yy}}+{\frac {\partial }{\partial x}} \left(
2\,{ \it h_{x}}\,W \right) +{\frac {\partial }{\partial y}} \left(
2\,{\it h_{y}}\,W \right) =0
\end{equation}

if condition
\begin{equation} \label{eq4}
{\it u}=-{\it h_{xx}}-{\it h_{yy}}+{{\it h_{x}}}^{2}+{{\it
h_{y}}}^{2}
\end{equation}
holds.

The Fokker-Planck equation \eqref{eq3} has the conservation low
form that yields a pair of potential equations

\begin{equation} \label{eq5}
{\it W_{x}}+ 2\,{ \it h_{x}}\,W  -{\it Q_{y}} =0
\end{equation}
\begin{equation} \label{eq6}
{\it W_{y}}+ 2\,{ \it h_{y}}\,W  +{\it Q_{x}} =0
\end{equation}

Let us consider linear operator

\begin{equation*}
\hat L\left( h \left( x ,y \right)  \right) \, {\bf F}=
\begin{pmatrix} { 2\,{ \it h_{x}}+\frac {\partial }{\partial x} }  &
{ -\frac {\partial }{\partial y} } \\ {{2\,{ \it h_{y}}+\frac
{\partial }{\partial y}}}  & {\frac {\partial }{\partial x}}
\end{pmatrix}\,
\begin{pmatrix} F_1 \\ F_2 \end{pmatrix}
\end{equation*}

Consider Darboux transformation in the form

\begin{equation*}
\hat L_D \, {\bf F}=
\begin{pmatrix} { r_{11}-a_{11}\,\frac {\partial }{\partial x}-b_{11}\,\frac {\partial }{\partial y}  }  &
{  r_{12}-a_{12}\,\frac {\partial }{\partial x}-b_{12}\,\frac
{\partial }{\partial y} } \\ { r_{21}-a_{21}\,\frac {\partial
}{\partial x}-b_{21}\,\frac {\partial }{\partial y} }  & {
r_{22}-a_{22}\,\frac {\partial }{\partial x}-b_{22}\,\frac
{\partial }{\partial y} }
\end{pmatrix} \,
\begin{pmatrix} F_1 \\ F_2 \end{pmatrix}
\end{equation*}

If linear operators $\hat L$ and $\hat L_{D}$ hold the relation

\begin{equation} \label{eq7}
\left( \hat L\left( h \left( x ,y \right) + \delta h \left( x ,y
\right) \right)\hat L_{D} - \hat L_{D} \hat L\left( h \left( x ,y
\right) \right) \right) \, {\bf F}= 0
\end{equation}

for any $ {\bf F} \in  \mathcal{F} \supset Ker\left( \hat L\left(
h \right)\right)$ where  $Ker\left( \hat L\left( h
\right)\right)=\{{\bf F}:{\hat {L}}\left( h \right){\bf F}=0\}$,
then for any ${\bf F_s}\in Ker\left( \hat L\left( h
\right)\right)$
 the function $\tilde {\bf F} \left( t,x \right)=
\hat L_{D} {\bf F_s} \left( t,x \right)$ is a solution of the
equation ${\hat {L}}\left( {\tilde {h}} \right) \tilde {\bf F} =0$
\, with new potential $\tilde h = h+\delta h$.

If one consider equation \eqref{eq7} on the set ${\mathcal{F}}$ of
arbitrary functions then treating $F_1, F_2$ and each of its
derivatives as independent variables the following equation can be
obtained

\begin{equation} \label{eq8}
{\left({\frac {\partial \left( h+\delta h \right)}{\partial x}}
\right)}^{2}+{\left({\frac {\partial \left( h+\delta h
\right)}{\partial x}} \right)}^{2}={\left({\frac {\partial h
}{\partial x}} \right)}^{2}+{\left({\frac {\partial  h }{\partial
x}} \right)}^{2}
\end{equation}

This is strong limitation for new potential $\tilde h = h+\delta
h$.

Let us consider equation \eqref{eq7} on the following set of
functions: ${\mathcal{F}}_0=\{{\bf F}: {F_1}_{x}+ 2\,{ \it
h_{x}}\, F_1  - {F_2}_{y} =0\}$. Taking into account this
dependance of $F_1, F_2$ derivatives the equations for $\delta h,
r_{ij}, a_{ij}, b_{ij}$ can be obtained. The particular solution
of this equations has the form

\begin{equation} \label{eq9}
\delta h = -\ln  \left( B \left( x,y \right)  \right)
\end{equation}

\begin{equation} \label{eq10}
\hat L_D =
\begin{pmatrix} { BR_{1}-B\frac {\partial }{\partial y}  }  &
{  BR_{2} } \\ { {B}_{x}-BR_{2} }  & { {B}_{y}+BR_{1} -B\frac
{\partial }{\partial y}}
\end{pmatrix} \,
\end{equation}

where $R_{1}, R_{2}$ are expressions in terms of $B, h$ and $B$
satisfy the system of two nonlinear differential equations.

Let us restrict further consideration by the simple case $h=0$. In
this case the expressions for $R_{1}, R_{2}$ have the form

\begin{equation} \label{eq11}
R_{1} = {\frac { {\it B_{y}}\, \left( {\it B_{xx}}-{\it B_{yy}}
\right) -2\,{\it B_{x}}\,{\it B_{xy}}  }{2\left({{\it
B_{x}}}^{2}+{{\it B_{y}}}^{2} \right)}}\,,\,\, R_{2} = {\frac
{{\it B_{x}}\, \left( {\it B_{xx}}-{\it B_{yy}} \right) +2\,{\it
B_{y}}\,{\it B_{xy}}  }{2\left({{\it B_{x}}}^{2}+{{\it
B_{y}}}^{2}\right)}}
\end{equation}

The system of equations for $B$ have the form
\begin{multline} \label{eq12}
 - \left( 2\,B{\it
B_y}\,{\it B_{xy}}+B{\it B_x}\, \left( {\it B_{xx}}-{\it B_{yy}}
 \right) +{\it B_x} \left( {{\it B_y}}^{2}+{{\it B_x}}^{2}
 \right)  \right)  \left( {\it B_{xx}}+{\it B_{yy}} \right)
\\
\shoveright{ +B \left( {{\it B_y}}^{2}+{{\it B_x}}^{2} \right)
{\frac {\partial }{
\partial x}} \left( {\it B_{xx}}+{\it B_{yy}} \right) =0}
\\
\shoveleft{ - \left( 2\,B{\it B_x}\,{\it B_{xy}}-B{\it B_y}\,
\left( {\it B_{xx}}-{\it B_{yy}}
 \right) +{\it B_y}\, \left( {{\it B_y}}^{2}+{{\it B_x}}^{2}
 \right)  \right)  \left( {\it B_{xx}}+{\it B_{yy}} \right)}
 \\
 +B \left( {{\it B_y}}^{2}+{{\it B_x}}^{2} \right) {\frac
{\partial }{
\partial y}} \left( {\it B_{xx}}+{\it B_{yy}} \right)
 =0
\end{multline}

The initial potential $u$ of the Schr\"odinger equation
corresponds to $h=0$ and according to equation \eqref{eq4} $u=0$.
The new potential of Schr\"odinger equation corresponds to $\delta
h$ and according to equations \eqref{eq4}, \eqref{eq9} is given by

\begin{equation} \label{eq13}
\tilde {u} \left( x ,y \right) = {\frac {\left( B_{xx}+B_{yy}
\right)}{B}}
\end{equation}

Note that according to formula  \eqref{eq13} $B$ is an example of
solution for the Schr\"odinger equation with potential $\tilde
{u}$. For $h=0$ initial system of potential equations \eqref{eq5},
\eqref{eq6} has the form

\begin{equation} \label{eq14}
W_{x}-Q_{y}=0, \,\, W_{y}+Q_{x}=0
\end{equation}

and according \eqref{eq10} we have for the solution of the new
Fokker-Planck equation with potential $\delta h$

\begin{equation} \label{eq15}
\tilde {W} \left( x ,y \right) = B\,R_{1}\,W-B \, \frac {\partial
W}{\partial y}  + B\,R_{2}\,Q
\end{equation}

where $W, Q$ are solutions of the system of equations
\eqref{eq14}.

The equations \eqref{eq2}, \eqref{eq9} provide for the solution of
the new Schr\"odinger equation with potential $\tilde {u}$ the
relation $\tilde {Y}=\tilde {W}/B$. The equation \eqref{eq2} in
the case $h=0$ provide for the initial solution $Y$ of the
Schr\"odinger equation with potential $u=0$ that $Y=W$. Then
according \eqref{eq15} the Darboux transformation for the
Schr\"odinger equation is

\begin{equation} \label{eq16}
\tilde {Y} \left( x ,y \right) = R_{1}\,Y-\frac {\partial
Y}{\partial y}  + R_{2}\,Q
\end{equation}

This is nonlocal Darboux transformation as the potential variable
$Q$ is a nonlocal variable connected with $Y$ by the system

\begin{equation} \label{eq17}
Y_{x}-Q_{y}=0, \,\, Y_{y}+Q_{x}=0
\end{equation}

To obtain the first example for the solution of equations
\eqref{eq12} let us consider ansatz $B=f(xy)$. This ansatz provide
in particular the solution

\begin{equation} \label{eq18}
B_s=\tanh \left( {\frac {xy-{\it C_2}}{{\it C_1}}} \right)
\end{equation}

 where ${\it C_1}, {\it C_2}$ are arbitrary constants. Note that
 $B_s$ and $1/B_s$ are not solutions of the Laplace equation. The
 solution $B_s$ provides by the formula \eqref{eq13}

\begin{equation} \label{eq19}
{\tilde {u}} =-2\, {{\it C_1}}^{-2}\left( {x}^{2}+{y}^{2} \right)
\left( \cosh \left( {\frac {xy-{\it C_2}}{{\it C_1}}} \right)
\right) ^{-2}
\end{equation}

\section{Decaying at infinity smooth rational potentials of Schr\"odinger equation }

The system of equations \eqref{eq12} for $B$ has some properties
that help to get its solutions. It can be proved by
straightforward calculation that if $B$ is a solution of equations
\eqref{eq12} then $1/B$ and $CB$ where $C$ is an arbitrary
constant are solutions as well. It is obvious from the form of
equations \eqref{eq12} that any solution of the Laplace equation
$B_{xx}+B_{yy}=0$ is the solution of these equations.

The solution $B_L$ of the Laplace equation provide $\tilde {u}=0$
according to the formula \eqref{eq13}. Taking $B=1/B_L$ one
obtains nontrivial $\tilde {u}$ but potentials of this kind have
singularities. To avoid singularities the special ansatz for $B$
can be used. Let us consider

\begin{equation} \label{eq20}
S_n=\sum\limits_{i=0}^n {\frac {{\it p_i}\, \left( x-{\it x_i}
\right) +{\it q_i}\, \left( y-{\it y_i} \right) }{ \left( x-{\it
x_i} \right) ^{2}+ \left( y-{ \it y_i} \right) ^{2}}}={\frac
{N_n}{M_n}}
\end{equation}

where ${\it p_i},{\it q_i},{\it x_i},{\it y_i}$ are arbitrary
constants, $N_n$ is the numerator of $S_n$ and the denominator
$M_n$ has the form

\begin{equation} \label{eq21}
M_n=\prod\limits_{i=0}^n  {\left(\left( x-{\it x_i} \right) ^{2}+
\left( y-{ \it y_i} \right) ^{2}\right)}
\end{equation}

The function $S_n$ is a solution of the Laplace equation. Let us
consider the following ansatz for $B$

\begin{equation} \label{eq22}
B_n={\frac {N_n}{M_n+C}}
\end{equation}

where $C$ is a constant. If $C>0$ then denominator of $B_n$ is
positive.

Substituting the ansatz \eqref{eq22} into equations \eqref{eq12}
it can be verified by straightforward calculation that $B_n$ is a
solution of equations \eqref{eq12} for small $n$ if $N_n$ is a
solution of the Laplace equation.

The first example of the rational solution is

\begin{equation} \label{eq23}
B_0={\frac {{\it p_0}\, \left( x-{\it x_0} \right) +{\it q_0}\,
\left( y-{\it y_0} \right) }{ \left( x-{\it x_0} \right) ^{2}+
\left( y-{ \it y_0} \right) ^{2}+C}}
\end{equation}

The solution $B_0$ provides by the formula \eqref{eq13}

\begin{equation} \label{eq24}
{\tilde {u}}_0 ={\frac {-8\,C}{ \left(  \left( x-{\it x_0} \right)
^{2}+ \left( y-{ \it y_0} \right) ^{2}+C \right) ^{2}}}
\end{equation}

The second example of the rational solution $B_1$ can be obtained
from the formula \eqref{eq22} where

\begin{multline} \label{eq25}
N_1=
\\
\left(  {\it p_0} \left({\it x_0} -{\it x_1} \right)- {\it q_0}
\left( { \it y_0}-{\it y_1} \right) \right)
\left({x}^{2}-{y}^{2}\right) +  2 \left(  {\it p_0}\left( {\it
y_0}-{\it y_1} \right)+{\it q_0}\left({\it x_0} -{\it x_1} \right)
\right) x y
\\
-\left({\it p_0}\left( {{\it x_0}}^{2}-{{\it x_1}}^{2}+{{\it y_0
}}^{2} -{{\it y_1}}^{2}\right) +  2\,{\it q_0} \left({\it
x_0}\,{\it y_1}-{\it y_0}\,{\it x_1} \right) \right) x
\\
+ \left( 2\,{\it p_0}\left( { \it x_0}\,{\it y_1}-{\it y_0}\,{\it
x_1} \right) - {\it q_0}\left( {{\it x_0}}^{2}-{{\it
x_1}}^{2}+{{\it y_0}}^{2}-{{\it y_1}}^{2} \right)  \right) y
\\
- {\it p_0}\left( {\it x_0}\left({{\it x_1}}^{2}+{{\it
y_1}}^{2}\right)-{\it x_1}\left({{\it x_0 }}^{2}+{{\it
y_0}}^{2}\right) \right) -{\it q_0} \left( {\it y_0}\left({{\it
x_1}}^{2}+{{\it y_1}}^{2}\right)-{\it y_1}\left({{\it x_0
}}^{2}+{{\it y_0}}^{2}\right) \right)
\end{multline}

The solution $B_1$ provides by the formula \eqref{eq13} the
potential

\begin{equation} \label{eq26}
{\tilde {u}}_1 ={\frac {-32\, C\left(  \left( x-1/2\,\left({\it
x_0}+{\it x_1}\right) \right) ^{2}+ \left( y-1/2\, \left({\it
y_0}+{\it y_1}\right) \right) ^{2}
 \right) }{ \left(  \left(  \left( x-{\it x_0} \right) ^{2}+ \left(
y-{\it y_0} \right) ^{2} \right)  \left(  \left( x-{\it x_1}
 \right) ^{2}+ \left( y-{\it y_1} \right) ^{2} \right) +C \right) ^{2
}}}
\end{equation}

In the case $ {\it x_0}=0,\,{\it y_0}=0,\,{\it x_1}=-8/17,\,{\it
y_1}=-2/17,\,C=160/17$ the potential ${\tilde {u}}_1$ coincides
with the first example of the Schr\"odinger equation potential
obtained in the papers \cite {Tsarev2007}, \cite {Tsarev2008} by
twofold application of the Moutard transformation.

Two arbitrary constants ${\it p_0},{\it q_0}\,\,$ in the formula
\eqref{eq25} yield two-parameter family of solutions $B$ for the
Schr\"odinger equation with potential ${\tilde {u}}_1$. Other
examples of solutions for the Schr\"odinger equation with
potential ${\tilde {u}}_1$ can be obtained by the formula
\eqref{eq16}.

For $n=2$ the numerator $N_2$ for the solution $B_2$ of the
equations \eqref{eq12} can be obtained from the formula
\eqref{eq20} and the condition that $N_2$ is a solution of the
Laplace equation. As for $n=1$ this yields two-parameter family of
solutions $B_2$. To avoid lengthy formulae we omit the expression
for $B_2$ and give the result for the potential in the case ${\it
x_0}=0,\,{\it y_0}=0$

\begin{equation} \label{eq27}
\begin{split}
{\tilde {u}}_2 &= {\frac {-8\,C\,G\left( x,y \right) }{ \left(
\left( {x}^{2}+{y}^{2} \right)  \left(
 \left( x-{\it x_1} \right) ^{2}+ \left( y-{\it y_1} \right) ^{2}
 \right)  \left(  \left( x-{\it x_2} \right) ^{2}+ \left( y-{\it
 y_2
} \right) ^{2} \right) +C \right) ^{2}}}
\\
G\left( x,y \right) &=  \left(  \left( 3\,x-2\,{\it k_1}
 \right) ^{2}+ \left( 3\,y-2\,{\it k_2} \right) ^{2} \right)\left( {x}^{2}+{y}^{2} \right)
  +6\,
 \left( {\it k_3}-{\it k_4} \right)  \left( {x}^{2}-{y}^{2} \right)
\\
& +12\, \left( {\it k_5}+{\it k_6} \right) x y -4\, \left( {\it
k_1}\,{ \it k_3}+{\it k_5}\,{\it y_2}+{\it k_6}\,{\it y_1} \right)
x
\\
& -4\, \left( {\it k_2}\,{\it k_4}+{\it k_5}\,{\it x_1}+{\it
k_6}\,{ \it x_2} \right) y+ \left( {{\it x_1}}^{2}+{{\it y_1}}^{2}
\right)
 \left( {{\it x_2}}^{2}+{{\it y_2}}^{2} \right)
\end{split}
\end{equation}
where ${{\it k_1}=\it x_1}+{\it x_2},\,{\it k_2}={\it y_1}+{\it
y_2},\,{\it k_3}={\it x_1}\,{\it x_2},\,{\it k_4}={\it y_1}\,{\it
y_2},\,{\it k_5}={\it x_1}\,{\it y_2},\,{\it k_6}={\it y_1}\,{\it
x_2}\,$.

In the case ${\it x_1}=-{\frac {1}{80}}-{\frac {1}{80}}\,\sqrt
{788+ \sqrt {1252969}}$,\,${\it y_1}={\frac {-159+\sqrt {788+\sqrt
{ 1252969}}}{16\,\sqrt {788+\sqrt {1252969}}}}$,\,${\it
x_2}=-{\frac {1}{80}}+{\frac {1}{80}}\,\sqrt { 788+\sqrt
{1252969}}$,\,${\it y_2}={\frac {159+ \sqrt {788+\sqrt
{1252969}}}{16\,\sqrt {788+\sqrt {1252969}}}}$,\,$C=50$ the
potential ${\tilde {u}}_2$ coincides with the second example of
the Schr\"odinger equation potential obtained in the papers \cite
{Tsarev2007}, \cite {Tsarev2008} by twofold application of the
Moutard transformation.

For $n=3$ consider the simple case $ {\it x_0}={\it x_2}={\it
x_3}=0,\,{\it y_0}={\it y_2}={\it y_3}=0$. In this case the
solution $B$ has the form

\begin{equation} \label{eq28}
\begin{split}
B_3 &={\frac {H\left( x,y \right)}{ \left( {x}^{2}+{y}^{2} \right)
^{3} \left(
 \left( x-{\it x_1} \right) ^{2}+ \left( y-{\it y_1} \right) ^{2}
 \right) +C}}
 \\
H\left( x,y \right) &= \left( {\it m_1}\,{\it p_1}+{\it m_2}\,{\it
q_1} \right)  \left(
 \left( {x}^{2}-{y}^{2} \right) ^{2}-4\,{x}^{2}{y}^{2} \right)
\\
  & +4\, \left( {\it m_1}\,{\it q_1}-{\it m_2}\,{\it p_1} \right) x y \left( {x
}^{2}-{y}^{2} \right)
 + \left( {\it m_3}\,{\it q_1}-{\it
m_4}\,{\it p_1} \right) x \left( {x}^{2}-3\,{y}^{2} \right)
\\
& + \left( {\it m_4}\,{ \it q_1}+{\it m_3}\,{\it p_1} \right) y
\left( {y}^{2}-3\,{x}^{2}
 \right)
\end{split}
\end{equation}
where ${\it m_1}={\it x_1}\left( {{\it x_1}}^{2}-3\,{{\it
y_1}}^{2} \right) ,\,{\it m_2}={\it y_1}\left({{\it y_1}}^{2}
-3\,{{\it x_1}}^{2} \right) ,\,{\it m_3}=2\,{\it x_1}{\it y_1}
\left( {{\it x_1}}^{2}+{{\it y_1}}^{2}
 \right)$, ${\it m_4}={{\it x_1}}^{4}-{{\it y_1}}^{4}\,$.

The solution $B_3$ provides by the formula \eqref{eq13} the
potential

\begin{equation} \label{eq29}
{\tilde {u}}_3 ={\frac {-128\,C \left( {x}^{2}+{y}^{2} \right)
^{2} \left(  \left( x-3 /4\,{\it x_1} \right) ^{2}+ \left(
y-3/4\,{\it y_1} \right) ^{2}
 \right) }{ \left(  \left( {x}^{2}+{y}^{2} \right) ^{3} \left(
 \left( x-{\it x_1} \right) ^{2}+ \left( y-{\it y_1} \right) ^{2}
 \right) +C \right) ^{2}}}
\end{equation}

Two arbitrary constants ${\it p_1},{\it q_1}\,\,$ in the formula
\eqref{eq28} yield two-parameter family of solutions $B$ for the
Schr\"odinger equation with potential ${\tilde {u}}_3$. Other
examples of solutions for the Schr\"odinger equation with
potential ${\tilde {u}}_3$ can be obtained by the formula
\eqref{eq16}.

\section{Results and Discussion}

The nonlocal Darboux transformation for the two - dimensional
stationary Schr\"odinger equation is obtained. It is shown that
this nonlocal transformation provides a useful tool for obtaining
exactly solvable two - dimensional stationary Schr\"odinger
operators. The examples of exactly solvable two - dimensional
stationary Schr\"odinger operators with smooth rational potentials
decaying at infinity are obtained. The values of the arbitrary
constants of rational potentials are indicated which provide the
examples of solvable Schr\"odinger operators obtained in the
papers \cite {Tsarev2007}, \cite {Tsarev2008} of I.A. Taimanov and
S.P. Tsarev.

\end{document}